\documentclass{elsart}
\journal{Physics Letter B}
\usepackage{epsfig}
\usepackage{bm}
\usepackage{graphicx}
\newcommand{\ra}{\rightarrow}
\newcommand{\sss}{s\bar{s}s}
\newcommand{\suu}{s\bar{u}u}
\newcommand{\uus}{u\bar{u}s}
\newcommand{\ccs}{c\bar{c}s}
\newcommand{\sdd}{s\bar{d}d}
\newcommand{\btosss}{b \ra \sss}
\newcommand{\btosssss}{b \ra \sss\bar{s}s}
\newcommand{\btouus}{b \ra \uus}
\newcommand{\btoccs}{b \ra \ccs}
\newcommand{\acp}{\mathcal{A}_{CP}}

\newcommand{\acppar}{\mathcal{A}_{CP}^0}
\newcommand{\dacppar}{\delta\acppar}

\newcommand{\bz}{B^0}
\newcommand{\bp}{B^+}
\newcommand{\bminus}{B^-}
\newcommand{\bpm}{B^{\pm}}
\newcommand{\nb}{{\rm N}_B}
\newcommand{\etac}{\eta_c}
\newcommand{\ks}{K_S^0}
\newcommand{\xs}{X_s}
\newcommand{\xsp}{X_s^+}

\newcommand{\xspm}{X_s^{\pm}}

\newcommand{\etackpm}{\etac K^{\pm}}

\newcommand{\bpmtoetackpm}{\bpm \to \etackpm}
\newcommand{\bztoetackstarz}{\bz \to \etac K^{*0}}
\newcommand{\bptoetackstarp}{\bp \to \etac K^{*+}}
\newcommand{\bpmtoetacxspm}{\bpm \to \etac \xspm}
\newcommand{\phiphixs}{\phi \phi \xs}
\newcommand{\phiphixsp}{\phi \phi \xsp}

\newcommand{\phiphixspm}{\phi \phi \xspm}
\newcommand{\phiphik}{\phi \phi K}

\newcommand{\phiphikpm}{\phi \phi K^{\pm}}
\newcommand{\bpmtophiphikpm}{\bpm \to \phiphikpm}
\newcommand{\btophiphixs}{B \to \phiphixs}
\newcommand{\bptophiphixsp}{\bp \to \phiphixsp}

\newcommand{\bpmtophiphixspm}{\bpm \to \phiphixspm}
\newcommand{\btophiphik}{B \to \phiphik}
\newcommand{\tnp}{\Theta_{\rm NP}}
\newcommand{\sintnp}{\sin \tnp}
\newcommand{\costnp}{\cos \tnp}
\newcommand{\ad}{a_{\rm D}}
\newcommand{\adtwo}{a_{\rm D}^2}
\newcommand{\ar}{a_{\rm R}}
\newcommand{\artwo}{a_{\rm R}^2}
\newcommand{\anp}{a_{\rm NP}}
\newcommand{\anptwo}{a_{\rm NP}^2}
\newcommand{\rtwo}{r^2}
\newcommand{\dsm}{D_{\rm SM}}
\newcommand{\dpmnp}{D^{\pm}_{\rm NP}}
\newcommand{\dpm}{D^{\pm}}
\newcommand{\bnp}{\mathcal{B}_{\rm NP}}
\newcommand{\bztophiks}{\bz \to \phi\ks}
\newcommand{\cals}{\mathcal{S}}
\newcommand{\cala}{\mathcal{A}}

\newcommand{\dmd}{\Delta m_d}

\begin{document}

%%% Comment out the following two lines 
%%% for PLB submission
\flushright{KEK Preprint 2003-107}
\vskip 3cm
%%%

\begin{frontmatter}

\title{Large Direct {\boldmath $CP$} Violation in {\boldmath $B$}
$\ra$ {\boldmath $\phi \phi {X_s}$} Decays}

\author{Masashi Hazumi}
%\email{masashi.hazumi@kek.jp}
\address{%
Institute of Particle and Nuclear Studies,\\
High Energy Accelerator Research Organization (KEK),\\
1-1 Oho, Tsukuba, Ibaraki 305-0801, Japan}
%\homepage{http://belle.kek.jp/~hazumi/}

%\date{\today}% It is always \today, today,
             %  but any date may be explicitly specified

\date{\today}

\begin{abstract}
We present a novel method to search for a new $CP$-violating
phase in the $\btosssss$ transition using $\btophiphixs$ decays,
where $\xs$ represents a final state with a specific strange
flavor such as $K^{\pm}$, $K^{*\pm}$ or $K^{\pm}\pi^{\mp}$.
Direct $CP$ violation can be enhanced 
due to an interference between an amplitude beyond the 
standard model (SM)
and the SM decay amplitude through the $\etac$ resonance.
We find that the $CP$ asymmetry can be as large as 0.4 
within the present experimental bounds on the $\btosss$ transition.
These decays provide a very clean experimental signature
and the background is expected to be small in particular
at $e^+e^-$ $B$ factories.
A simulation study for the $\bpmtophiphikpm$ decay
shows that the statistical significance of $CP$ violation
can exceed 5 standard deviations with $10^9$ $B$ mesons.
%a statistical error on the $CP$ asymmetry 
%will be $\sim$0.06 for $10^9$ $B$ mesons.
%its sensitivity to a new $CP$-violating phase is comparable 
%to that in mixing-induced $CP$ violation in $\bztophiks$ decays.
\end{abstract}


\end{frontmatter}

\clearpage

%%%%%%%%%%%%%%%%%%%%%%%%%%%%
%\section{Introduction}               % Introduction goes below.
The phenomenon of $CP$ violation is one of the major
unresolved issues in our understanding of elementary particles today.
In the standard model (SM),
$CP$ violation arises from an irreducible complex phase,
the Kobayashi-Maskawa (KM) phase~\cite{KM},
in the Cabibbo-Kobayashi-Maskawa (CKM) weak-interaction quark mixing 
matrix~\cite{KM,CABIBBO63}.
Recent observations of direct $CP$ violation in the neutral
$K$ meson system~\cite{KTeV,NA48} and mixing-induced
$CP$ violation in the neutral $B$ meson system~\cite{CP1_Belle,CP1_BaBar}
strongly support the KM mechanism.
Despite this success, additional $CP$-violating phases
are inevitable
in most of theories involving new physics (NP) beyond 
the SM~\cite{BSM}.
Some of them allow
large deviations from the SM predictions
in $B$ meson decays. Examples of such theories
include supersymmetric grand-unified theories
with the see-saw mechanism that can accommodate the large neutrino 
mixing~\cite{BTOSSS}.
Therefore it is of fundamental importance to 
measure $CP$ asymmetries that are sensitive to the difference
between the SM and NP.
Additional sources of $CP$ violation are also highly desirable
to understand the origin of the matter-antimatter asymmetry of
the universe;
detailed studies have found no way that $CP$ violation
in the SM alone could explain the baryogenesis~\cite{BARYOGENESIS}.
Many methods to search for a new source of $CP$ violation
in $B$ meson decays
have been proposed until now. One of the most promising ways
among them is to compare the mixing-induced $CP$ asymmetries 
in the $B \to \phi \ks$ decay~\cite{WORAH97}, which is dominated by 
the $\btosss$ transition that is known to be sensitive
to possible NP effects, 
with those in the $\bz \to J/\psi \ks$ decay~\cite{Sanda}.
Recent measurements by Belle~\cite{Belle_sss} and
BaBar~\cite{BaBar_sss} collaborations
yield values smaller than the SM expectation;
a difference by 2.6 standard deviations 
is obtained when two results are combined.
The other charmless decays $\bz \to \eta' \ks$ and
$\bz \to K^+K^-\ks$, which are
mediated by $\btosss$, $\suu$ and $\sdd$ transitions,
also provide additional information~\cite{Belle_sss,BaBar_sss}.
The present world average with $\bztophiks$, $\eta' \ks$
and $K^+K^-\ks$ combined is different from the average
with $\bz \to J/\psi \ks$ and related modes
by 3.1 standard deviations~\cite{HFAG}.
Possible implications of these
measurements are discussed in the literature~\cite{PHIKS}.

%%%%%%%%%%%%%%%%%%%%%%%%%%%%%%%%%%%%%%%%%%%%
%\section{Principle of the Resonance Method}
In this Letter, we propose a novel method to search for
a new $CP$-violating phase in the hadronic $b \to s$ transition
using $\bpmtophiphixspm$ decays~\cite{HAZUMI01}.
Here $\xspm$ represents a final state with a specific strange flavor
such as $K^{\pm}$ or $K^{*\pm}$.
These non-resonant direct decay amplitudes are dominated by
the $\btosssss$ transition.
A contribution from the $\btouus$ transition followed by rescattering
into $s\bar{s}s$ is expected to be small because of the 
CKM suppression and
the OZI rule~\cite{OZI}.
% and are not affected by
%a pollution from $\btouus$ tree contributions.
In these decays, when the invariant mass of the $\phi\phi$ system
is within the $\etac$ resonance region, they interfere with
the $\bpm \to \etac (\to \phi \phi) \xspm$ decay
that is dominated by the $\btoccs$ transition.
The decay width of
$\etac$ is sufficiently large~\cite{PDG02,Belle_etack}
to provide a sizable interference.
Within the SM, this interference does not cause sizable direct $CP$ violation
because there is no weak phase difference between the $\btosssss$ 
and the $\btoccs$ transitions.
On the other hand, a NP contribution with a new $CP$-violating phase
can create a large weak phase difference.
Thus large $CP$ asymmetries can appear only from NP amplitudes, and
an observation of direct $CP$ violation in these decays is an unambiguous
manifestation of physics beyond the SM.

Although the same argument so far is applicable 
to the $\bpm \to \phi \xspm$ decays,
there is no guaranteed strong phase difference that is calculable
reliably for these decays.
In contrast,
the Breit-Wigner resonance provides 
the maximal strong phase difference in the case of
$\bpm \to (\phi \phi)_{m \sim m_{\etac}} \xspm$ decays.
$CP$ violation that arises from
interference between a resonant and a nonresonant amplitude
was initially studied in top-quark decays~\cite{EILAM91}
and in radiative $\bpm$ decays~\cite{ATWOOD94-95}.
The charmonium width effect
on the SM direct $CP$ asymmetries in $B$ decays was 
first considered in
$\bpm \to \etac (\chi_{c0})\pi^{\pm}$ decays~\cite{EILAM95} and
was also used in subsequent theoretical studies~\cite{EILAM95-2}.

Experimentally, an evidence for the decay $\bpmtoetackpm$ was first reported
by the CLEO collaboration~\cite{CLEO00-12}.
%where one of the important $\etac$ decay modes
%is $\etac \to \phi (\to K^+K^-) \phi (\to K^+K^-)$.
%Thanks to a very distinctive final state with five charged
%kaons together with the $\phi$ mass constraint,
%the estimated background level was reasonably low. 
Recently Belle~\cite{Belle_etack} has also reported 
the first observation of the $\bztoetackstarz$ decay.
This implies that other modes such as 
$\bptoetackstarp$ will also be seen with a similar branching
fraction, so that we will be able to study
semi-inclusive $\bpmtoetacxspm$ transitions experimentally.
The semi-inclusive branching fraction of $\bpmtoetacxspm$ is not yet measured,
but is theoretically expected
to be comparable to the branching fraction of the semi-inclusive
decay $\bpm \to J/\psi \xspm$~\cite{ETACBR}. 

%%%%%%%%%%%%%%%%%%%%%%%%%%%%%%%
%\section{General Formalism}
%\label{sec:formalism}
%\newpage

We derive the rates and the asymmetry of
the decays $\bpm \to (\phi \phi)_{m \sim m_{\etac}} \xspm$
based on the formalism described
in the study of $\bpm \to \etac (\chi_{c0})\pi^{\pm}$ decays~\cite{EILAM95}.
The distribution of two $\phi$'s
is determined with two kinematical variables;
one is the invariant mass of the $\phi \phi$ system, $m$, and
the other is the angle $\theta$ between the $B$-meson momentum and the momentum
of one of two $\phi$'s in the center-of-mass frame of the $\phi \phi$ 
system.
To have the interference between resonant and direct
amplitudes, $m$ should be in the $\etac$ resonance
region. To be specific, we require in this study that the difference
between $m$ and $\etac$ mass ($M$) should satisfy
$|m-M| < 3\Gamma$, where $\Gamma$ is the
the width of the $\etac$ resonance~\cite{footnoteETAC}. 
The differential decay rate normalized with the total $\bpm$ decay rate
is then given by the following equation:
\begin{equation}
\frac{1}{\Gamma_B}\frac{d\Gamma^{\pm}}{dz}=
\int_{(M-3\Gamma)^2}^{(M+3\Gamma)^2} ds \vert R(s)+D^{\pm}(s, z)\vert^2~,
\end{equation}
where $R(s)$ is the resonant amplitude, 
$D^{\pm}(s, z)$ is the direct amplitude of the 
$\bpmtophiphixspm$ decay, $\Gamma_B$ is the total $B$ decay
rate, $s \equiv m^2$ and $z \equiv \cos \theta$.

The resonant amplitude $R(s)$ is given by 
\begin{equation}
R(s)\equiv A(\bpmtoetacxspm \to \phiphixspm)=
\frac{\ar\sqrt{M \Gamma}}{(s-M^2)+iM\Gamma}~,
\end{equation}
where $\ar$ is a product of the
weak decay amplitude of $\bpmtoetacxspm$
and the real part of the $\etac$ decay amplitude to $\phi\phi$.

The direct amplitude $\dpm$
is separated into contributions
from the SM, $\dsm$, and from NP, $\dpmnp$,
\begin{eqnarray}
\dpm(s\approx M^2, z) & \equiv & \dsm(s\approx M^2, z)\nonumber\\ 
                         & +      & \dpmnp(s\approx M^2, z),\\
\dsm(s\approx M^2, z) & \equiv & \frac{\ad(z)}{\sqrt{M \Gamma}}~e^{i\delta},\\
\dpmnp(s\approx M^2, z) & \equiv & 
\frac{\anp(z)}{\sqrt{M \Gamma}}~e^{i\delta'}e^{{\pm}i\tnp},
\end{eqnarray}
where $\ad(z)$ is a real part of the SM direct amplitude,
$\delta$ ($\delta'$) is a strong phase difference between the resonant amplitude
and the SM (NP) direct amplitude,
$\anp(z)$ is a real part of the NP amplitude
and $\tnp$ is a new $CP$-violating phase.
If $\delta \neq \delta'$ holds, direct $CP$ violation can also occur
from an interference between the SM and NP direct amplitudes.
We do not take this case in our study and assume $\delta = \delta'$
in the following discussion.

The difference between the decay rates of $\bp$ and $\bminus$ is 
given by
\begin{equation}
\label{eq:dgp-dgm}
{1\over \Gamma_B}({d\Gamma^{+}\over dz}-{d\Gamma^{-}\over dz})
\equiv \gamma^-(z)
\cong -4\pi \ar \anp(z)\cos \delta \cdot \sintnp~.
\end{equation}
%where we used a close approximation in integrating the variable $s$.
Similarly the sum of two decay rates is given by
%\begin{widetext}
\begin{eqnarray}
\label{eq:dgp+dgm}
{1\over \Gamma_B}({d\Gamma^{+}\over dz}+{d\Gamma^{-}\over dz})
\equiv \gamma^+(z)
\cong
2\pi \artwo + 24 \adtwo(z)(\rtwo+2r\costnp + 1)
\mbox{} \nonumber\\
- 4\pi \ar \ad(z)(r\costnp + 1) \sin \delta,
\end{eqnarray}
%\end{widetext}
where 
$r \equiv \anp(z)/\ad(z)$ is the amplitude ratio of NP
to the SM.
The $z$ dependence of $r$ reflects
the spin components of the $\phi\phi$ system, which can be determined
at $B$ factories in the future
from the differential decay rates in the mass-sideband region below
the $\etac$ resonance.
Although only a pseudo-scaler component in the direct transition
interferes with the $\etac$ resonance, 
the effect of other components can be estimated
by such a measurement.
Thus, for simplicity, we assume that the direct transition
is dominated by a pseudo-scaler component and
ignore the $z$ dependence of $r$ in the following discussion.

The maximum asymmetry is realized when 
$\cos \delta \simeq 1$
is satisfied. Assuming that $\delta$ is small following the
discussion by Eilam, Gronau and Mendel~\cite{EILAM95},
the differential partial rate asymmetry is
\begin{equation}
\label{eq:Acp(z)}
A_{CP}(z) \equiv \frac{\gamma^-(z)}{\gamma^+(z)}
          \cong  \frac{-4\pi \ar \anp(z) \sintnp}
           {2\pi \artwo + 24 \adtwo(z)(\rtwo + 2r \costnp + 1)}~.
\end{equation}
As a measure of $CP$ violation, we define the following $CP$-asymmetry parameter:
\begin{equation}
\acp \equiv \sqrt{\frac{\int _{-1}^{1} dz \gamma^-(z)^2}{\int _{-1}^{1} dz \gamma^+(z)^2}}.
\end{equation}
%
%%%%%%%%%%%%%%%%%%%%%%%%%%%%%%%%%%%%
%\section{Estimation of $CP$ Asymmetry}
%\label{sec:estimation}
The numerator of $\acp$ can be expressed with the 
branching fraction of the resonance ($2\pi \artwo$) and that of
NP in the resonance region 
($\mathcal{B}_{\rm NP}$):
\begin{equation}
{\int _{-1}^1} dz \gamma^-(z)^2 = (2\pi \artwo) \cdot \mathcal{B}_{\rm NP}
                                       \cdot \frac{2\pi}{3} \sin^2 \tnp~,
\end{equation}
where
\begin{equation}
2\pi \artwo \cong \mathcal{B}(\bpmtoetacxspm) \cdot \mathcal{B}(\etac \to \phi\phi)~,
\end{equation}
and
\begin{equation}
\mathcal{B}_{\rm NP} \equiv {1 \over {M \Gamma}}{\int_{(M-3\Gamma)^2}^{(M+3\Gamma)^2}} ds 
                                                   {\int_{-1}^1} dz \anptwo(z).
\end{equation}
The inclusive branching fraction $\mathcal{B}(\bpmtoetacxspm)$ has not been
measured and only an upper limit of 0.9\% is available~\cite{PDG02}. 
Therefore we take theoretical expectations that range from
0.3\% to 0.7\%, depending on the estimation of
the ratio of $\mathcal{B}(B \to \etac X)$ to $\mathcal{B}(B \to J/\psi X)$~\cite{ETACBR}.
For $\mathcal{B}(\etac \to \phi\phi)$, we take the present world average
of ($0.71\pm 0.28)$\%~\cite{PDG02}.
As a result we obtain $2\pi \artwo = (2\sim 5) \times 10^{-5}$.

There is no direct experimental bound on the $\btophiphixs$ decay
(see {\it Note added} in the end of this Letter). 
A bound on the $B \to \phi \xs$ decay ($2.2\times 10^{-4}$ at 90\%C.L.) 
is available~\cite{CLEO95}. The search, however,
was devoted only for energetic $\phi$ mesons above 
$p_\phi > 2.1$~GeV/$c$. On the other hand, 
the momentum of $\phi$ mesons 
in the $\btophiphixs$ decay is lower;
it should be less than 2.2 GeV/$c$.
Consideration on other experimental bounds also leads to
a weak limitation on $b \to s g$~\cite{KAGAN95};
$\mathcal{B}(\btosss) \sim 1$\% is still allowed, 
while the estimation of $\mathcal{B}(\btosss)$ within the SM is 
$\sim 0.2$\%~\cite{DESHPANDE96}.
Since there is no reliable way to calculate the hadronic matrix element of
the multi-body decay $\btophiphixs$,
we use an event generator that is based on the LUND fragmentation model~\cite{LUND}
for the estimation of  $\bnp$. 
We start from the $b \to sg^* \to s\bar{s}s$ transition
that mainly produces soft $\phi$ mesons.
We find that it is possible to assume $\mathcal{B}(b \to sg^* \to s\bar{s}s) \sim 1$\%
without conflicting the existing experimental results 
such as the branching ratio of the exclusive 
$B \to \phi K$ decay~\cite{PHIK}.
%The ratios for the $q\bar{q}$ popping are 
%assumed to be $u\bar{u} : d\bar{d} : s\bar{s} = 1:1:0.3$
%which are the default values of JETSET7.4~\cite{LUND} and are consistent with
We use the default value of the $s\bar{s}$ popping probability in JETSET7.4~\cite{LUND},
which is consistent with
the ratio of the branching fractions between
$B \to J/\psi K \phi$ and $B \to J/\psi K$~\cite{CLEO9908014}.
%A $\phi$ meson is produced if and only if an $s\bar{s}$ pair is created and is chosen to be
%hadronized. 
We estimate $\bnp$ to be $ \sim 5 \times 10^{-6}$ 
for $\mathcal{B}(b \to sg^* \to s\bar{s}s) = 1$\%.
Estimations of the corresponding branching fraction within the SM
typically yield smaller values than this estimation.
For example, the branching fraction of $B \to \phi \xs$ is expected to
be around $10^{-4}$. With an additional $s\bar{s}$ popping and with
the aforementioned simulation, we obtain 
$\mathcal{B}(\btophiphixs)_{\rm SM} \sim 3\times 10^{-7}$ in the 
$\etac$ mass region. 
Also from the expected $\mathcal{B}(\btosss)$ in the SM,
we obtain $\sim 9\times 10^{-7}$.
Thus $r^2 \sim 5$ holds for $\mathcal{B}(b \to sg^* \to s\bar{s}s) = 1$\%.
Note that, even in this case, an observation of non-resonant 
$\btophiphixs$ decays
in the $\etac$ mass sideband region alone can not establish physics
beyond the SM because of the difficulty in estimating 
branching fractions of multi-body hadronic decays.

The evaluation of the denominator of $\acp$ is straightforward with
the estimations mentioned above.
We obtain the following result for 
$\mathcal{B}(b \to sg^* \to s\bar{s}s) \sim 1\%$ and
$2\pi \artwo = 2 \times 10^{-5}$: 
%and $|\sintnp| \sim 1$: 
%\begin{widetext}
\begin{eqnarray}
& \acp \cong 
            \sqrt{
		\frac{\bnp}
		      {\mathcal{B}(\bpmtoetacxspm)
	                   \cdot \mathcal{B}(\etac \to \phi\phi) +
		        2(1 + 2r^{-1}\costnp + r^{-2})\bnp}
	         }\nonumber\\
&       \times 	\sqrt{\frac{\pi}{3}} 
	\cdot |\sintnp| \sim 0.40 \cdot |\sintnp|~.
\end{eqnarray} 
%\end{widetext}
A large $CP$ asymmetry of 0.4 is allowed.
The asymmetry is roughly proportional to $|r|$.
Therefore it can be sizable even with $r^2 < 1$;
for example, $\acp \sim$ 0.1 is allowed for $r^2 = 0.3$.

A $CP$ asymmetry proportional to $\sin \phi_3$
($\phi_3 = 59^\circ \pm 13^\circ$~\cite{PDG02})
may arise within the SM due to an
interference between the resonant amplitude and
a SM contribution from the $\btouus$ transition followed by rescattering
into $s\bar{s}s$. The effect is expected to be small
since the $\btouus$ decay amplitude is CKM-suppressed,
and the rescattering amplitude is also suppressed by the OZI rule.
From the measured branching fractions for
$B^+ \to K^+K^-K^+$ and $B^+ \to K^+K^-\pi^+$ together with the additional
$O(\lambda)$ suppression, where $\lambda = 0.2229\pm 0.0022$
is the sine of the Cabibbo angle~\cite{PDG02},
the fraction of the $\btouus$ transition in the $B^+ \to K^+K^-K^+$
decay amplitude is estimated to be $\sqrt{0.022\pm0.005}$
in Ref.~\cite{BelleKKK02}.
Using this information and with a modest assumption on
a rescattering amplitude ratio 
$\mathcal{A}(u\bar{u} \to \phi\phi)/\mathcal{A}(u\bar{u} \to K^+K^-) \leq 0.3$
from the OZI rule, we estimate the $CP$ asymmetry to be less than 0.9\%.
Thus we conclude that the $CP$ asymmetry within the SM
is most likely below 1\%.

\begin{figure}[htbp]
  \begin{center}
    \includegraphics[width=1.0\textwidth,clip]{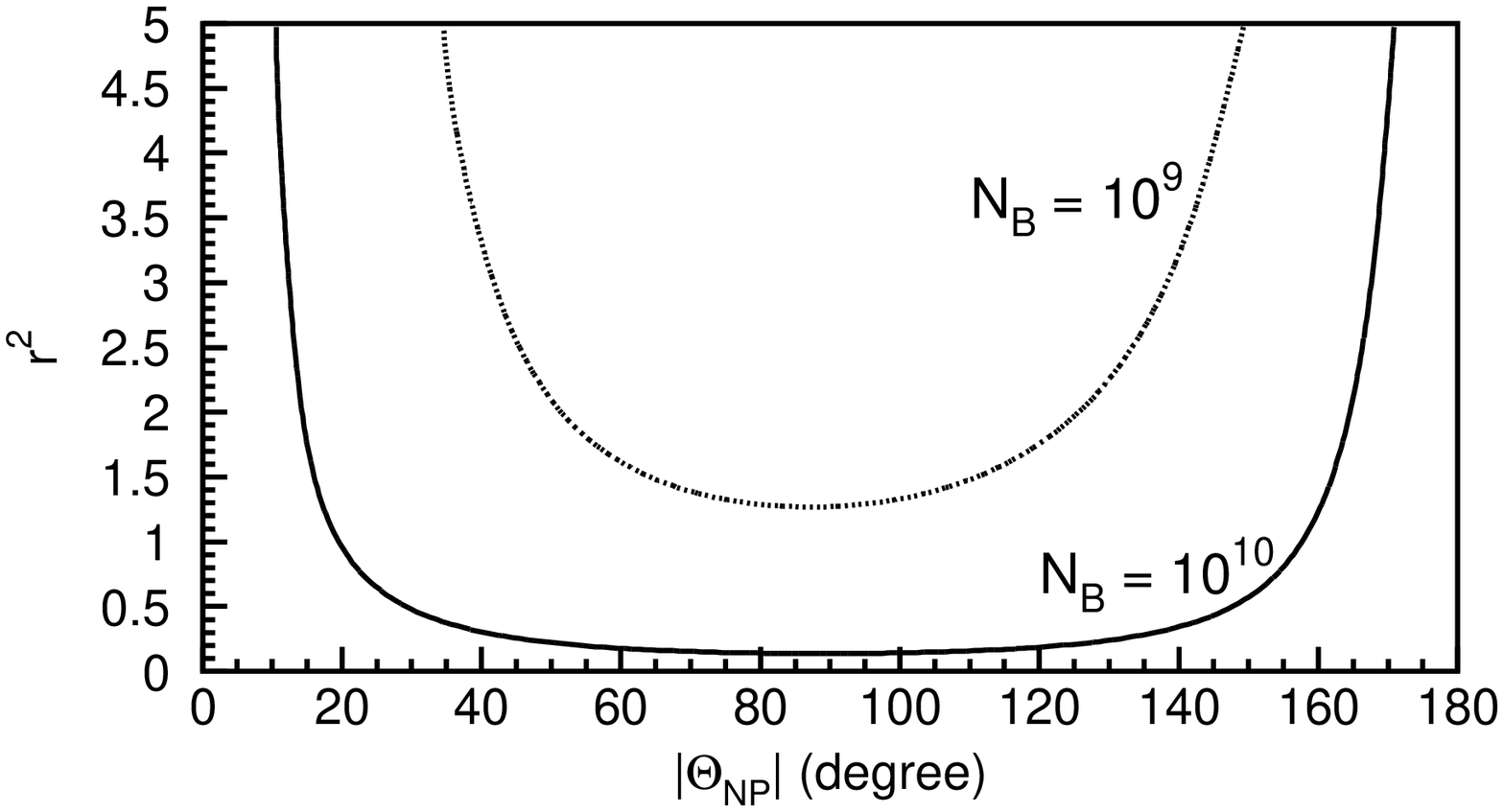}
  \end{center}
 \caption{Expected sensitivities on direct $CP$ violation in 
          the $\bpmtophiphikpm$ decay
          for $10^9$ $B$ mesons (dotted line) 
          and $10^{10}$ $B$ mesons (solid line).
          In the regions above the curves, direct $CP$ violation
          can be measured with a 5$\sigma$ significance or larger.}
\label{fig:tnpvsr2}
\end{figure}

%%%%%%%%%%%%%%%%%%%%%%%%%%%%%%%%%%%
%\section{Experimental Feasibility}
%\label{sec:feasibility}
%
We perform Monte Carlo simulation for
the $\bpmtophiphikpm$ decay and estimate statistical errors
on the $CP$ asymmetry parameter.
For this decay mode, the background
level is expected to be small enough to be neglected~\cite{HCHUANG}.
The reconstruction efficiency and the $\phi\phi$ mass resolution
are estimated using a GEANT-based detector simulator for
the Belle detector~\cite{Belle}.
We obtain $\sim$300 events for $\nb = 10^9$, where
$\nb$ is the number of charged $B$ mesons recorded by a detector.
We perform an unbinned maximum-likelihood fit
to the differential decay rate distribution, 
which is proportional to $\vert R(s)+D^{\pm}(s, z)\vert^2$,
instead of integrating the distribution.
We choose the following two free parameters in the fit:
$\acppar \equiv -2r(\ad/\ar)\sintnp$ and
$\mathcal{B} \equiv \adtwo(\rtwo+2r\costnp+1)$.
$\acppar$ is the $CP$ asymmetry in the Breit-Wigner term.
%and is close to $\acp$.
$\mathcal{B}$ is proportional to the branching ratio of the non-resonant 
$\bpmtophiphikpm$ decay below the $\etac$ mass region.
The statistical error for $\acppar$ is estimated to be $\dacppar \sim 0.06$.
Figure.~\ref{fig:tnpvsr2} shows the 5$\sigma$ search regions
for $\nb = 10^9$ (dotted line) and
for $\nb = 10^{10}$ (solid line), which will be accessible
at next-generation high-luminosity $e^+e^-$ $B$ factories.
Direct $CP$ violation will be observed in a large parameter space
above a 5$\sigma$ significance.

\begin{figure}[htbp]  
  \begin{center}
    \includegraphics[width=1.0\textwidth,clip]{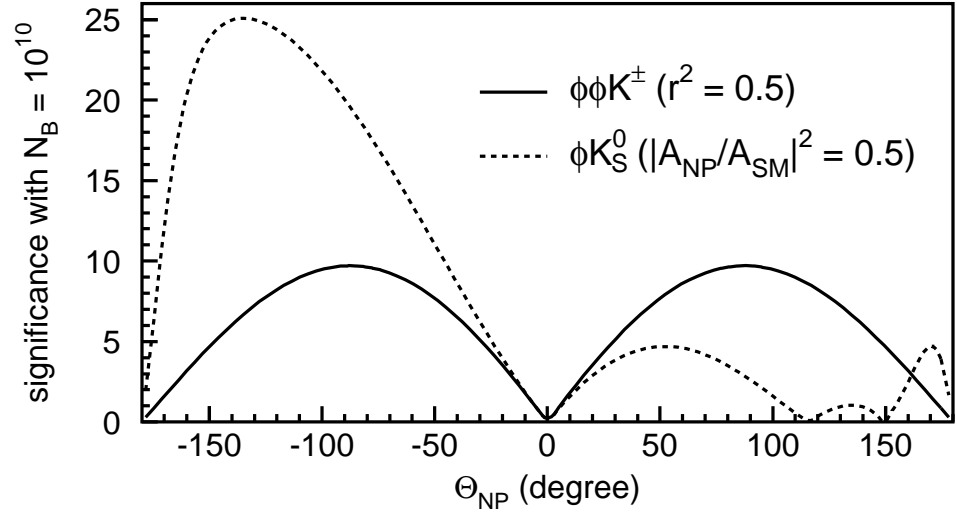}
  \end{center}
 \caption{Expected statistical significance of deviations from
          the SM for direct $CP$ violation in the $\bpmtophiphikpm$ decay
          with $\rtwo = 0.5$ (solid line)
          and for time-dependent $CP$ violation in
          the $\bztophiks$ decay with $|A_{\rm NP}/A_{\rm SM}|^2 = 0.5$
          (dashed line). For each case, significance is calculated
          with $10^{10}$ $B$ mesons.}
\label{fig:signp1010}
\end{figure}

%%%%%%%%%%%%%%%%%%%%%%%%%%%%%%%%%%%
%\section{Comparison between $\bpmtophiphikpm$ and $\bztophiks$}
%\label{sec:comparison}
%
The new $CP$-violating phase $\tnp$ also affects
time-dependent $CP$-violating asymmetries
$A_{CP}(t) = \cals \sin(\dmd t)+ \cala\cos(\dmd t)$
in $\bztophiks$ and related decays.
Here $\dmd$ is the mass difference 
between the two $B^0$ mass eigenstates, and
$\cals$ and $\cala$ are parameters for
mixing-induced $CP$ violation and direct $CP$ violation, respectively.
Ignoring a strong phase difference between 
the amplitude of NP ($A_{\rm NP}$) and SM ($A_{\rm SM}$),
we obtain
\begin{equation}
\cals = \frac{\sin2\phi_1+2\rho\sin(2\phi_1+\tnp)+\rho^2\sin(2\phi_1+2\tnp)}
              {1+\rho^2+2\rho\cos\tnp},
\end{equation}
where $\rho \equiv A_{\rm NP}/A_{\rm SM}$ is an amplitude ratio of NP to the SM
and $\phi_1$ is one of the angles of the unitarity triangle.
In particular, a difference in
$\cals$ between $\bztophiks$ and $\bz \to J/\psi \ks$ decays,
i.e. $\Delta \cals \equiv \cals(\phi\ks) - \cals(J/\psi\ks) \neq 0$,
would be a clear signal of the new phase since
$\cals(J/\psi\ks) = \sin2\phi_1$ is held to a good approximation.
We define expected statistical significance of the deviation from the SM 
by $\acppar/\dacppar$ for the $\bpmtophiphikpm$ decay and 
by $\Delta\cals/\delta\Delta\cals$
for the $\bztophiks$ decay, where
$\delta\Delta\cals$ is an expected statistical error
of $\Delta\cals$ extrapolated from the latest result
by the Belle experiment~\cite{Belle_sss}.
Although $\rtwo$ is not necessarily equal to
$\rho^2$, both decays are governed
by the same $\btosss$ transition. Therefore it is
reasonable to choose $\rtwo = \rho^2$ for comparison.
Figure.~\ref{fig:signp1010} shows the resulting
significance for $10^{10}$ $B$ mesons and with
$\rtwo = \rho^2 = 0.5$.
%The significance for $\acppar$
%is similar or superior to that for $\Delta \cals$
%in the region $\tnp > -50^\circ$. 
The significance for $\Delta \cals$
largely depends on the sign of $\tnp$, which
is not the case for the $\bpmtophiphikpm$ decay.
The sign dependence arises from an asymmetric
range for $\Delta \cals$; to a good approximation,
we have
$-1-\sin 2\phi_1 \leq \Delta\cals \leq 1-\sin 2\phi_1$
where $\sin 2\phi_1 = +0.736\pm 0.049$~\cite{HFAG}.
Therefore the $\bpmtophiphikpm$ decay plays a unique
role in searching for a new $CP$-violating phase.

%%%%%%%%%%%%%%%%%%%%%%%%%%%%%%%%%%%
%\section{Systematic uncertainties}
%\label{sec:syserror}
%
In the above estimation, we use parameters that have
uncertainties. However, they can in principle be
measured precisely if a sufficient number of 
$B$ mesons are produced. 
In our estimation, we assume efficiencies and background levels
that have been achieved with the Belle detector at the KEK $B$ factory.
They depend on the actual detector performance 
and beam conditions, which might be different
at a next-generation $B$ factory with the higher luminosity.
Detailed simulation studies as well as
some extrapolation from data at current $B$ factories
will be needed for further quantitative evaluation.

%%%%%%%%%%%%%%%%%%%%%%%%%%%%%%%%%%%
%\section{Discussion}
%\label{sec:discussion}
%
Experimental sensitivities can be improved by
adding more final states. The
technique to reconstruct $\xs$,
which has been successfully adopted for the measurements
of semi-inclusive
$B \to \xs \ell\ell$ transitions~\cite{KANEKO03},
can be used for this purpose.
Flavor-specific neutral $B$ meson decays,
such as $\bz \to \phi\phi K^{*0}(\to K^+\pi^-)$,
and other charmonia such as
the $\chi_{c0} \to \phi\phi$ decay can also
be included.
The method proposed in this Letter is also applicable to other 
beauty hadrons. Examples include
$B_s \to \etac (\to \phi \phi) \phi$ and 
$\Lambda_b \to \etac (\to \phi \phi) \Lambda$.

%%%%%%%%%%%%%%%%%%%%%%%%%%%%%%
%\section{Conclusion}
%\label{sec:conclusion}
In summary, we have investigated 
$\btophiphixs$ decays
as a probe for a new $CP$-violating phase
in the $\btosssss$ transition.
Large $CP$ violation may arise from
an interference between $\bp \to \etac (\to \phi\phi) \xsp$
and a contribution of the physics beyond the
SM to the non-resonant decay $\bptophiphixsp$.
We find that the direct $CP$ asymmetry 
can be as large as 0.4.
The experimental signature will be very clean
in particular at $e^+e^-$ $B$ factories;
when $10^9$ charged $B$ mesons are available,
the statistical error on the $CP$ asymmetry parameter
is estimated to be $\sim$0.06. 
The statistical significance of $CP$ violation
in the $\bpmtophiphikpm$ decay can
exceed 5 standard deviations. 
An even better sensitivity will be expected by including
other final states.
We also find that this result is comparable to the expected sensitivity
with mixing-induced $CP$ violation in the $\bztophiks$ decay.
Thus $\btophiphixs$ decays will be useful to search for
$CP$ violation beyond the SM in the $\btosssss$ transition
at high-luminosity $e^+e^-$ $B$ factories.

%%%%%%%%%%%%%%%%%%%%%%%%%%%%%%%%%%%
\section*{Acknowledgements}
\label{sec:acknowledgment}

I am grateful to I.~Bigi, T.~E.~Browder, W.~S.~Hou, Y.~Okada,
A.~I.~Sanda and A.~Soni for useful comments.
I would also like to thank H.-C.~Huang and T.~Hojo
for providing experimental information, and 
K.~Sumisawa for his help in Monte Carlo simulation and fitting.
This work was supported in part by 
a JSPS Grant-in-Aid for Scientific Research (C), 12640268.

{\it Note added.--}After finishing this study
and as we were preparing the manuscript,
the Belle Collaboration announced evidence for
$\btophiphik$ decays~\cite{HCHUANG03}.
The signal purity is close to 100\% 
when the $\phi\phi$ invariant mass is within the $\etac$ mass region,
as assumed in our study. The results are still limited by
small statistics; using the measured branching fractions of the
$\btophiphik$ decay within and below the $\etac$ mass region, we find
that the maximal $CP$ asymmetry of $\acp \sim 0.4$ is still
allowed. The same paper also reports 
$\mathcal{B}(\etac \to \phi\phi) = 
(1.8^{+0.8}_{-0.6}\pm0.7)\times 10^{-3}$,
which is smaller than the current world average. 
We repeat the fit procedure described in this Letter
with the above branching fraction.
Although the smaller value for $\mathcal{B}(\etac \to \phi\phi)$ 
results in the smaller number of signal events,
the $CP$ asymmetry from
the interference between the resonant and the new-physics
amplitudes becomes larger. We find that the change in 
$\mathcal{B}(\etac \to \phi\phi)$ does not largely affect
the significance; the difference is less than 10\%
for $\rtwo = 0.5$ and $\sintnp = 1$.

%%%%%%%%%%%%%%%%%%%%%%%%%%%%%%%%%%%

\clearpage
\newpage
%%%%%%%%%%%%%%%%%%%%
%%%%%%%%%%%%
%%%%%%%%%%%%%%%%%%%%
%%%%%%%%%%%%


\begin{thebibliography}{99}
\bibitem{footnoteETAC}
In this study we take $M = 2980$ MeV/$c^2$ and
$\Gamma = 29$ MeV,
which are the values from the recent measurements
by the Belle collaboration~\cite{Belle_etack}.
\end{thebibliography}
\end{document}